\documentclass{article}
\usepackage{graphicx}
\usepackage{latexsym}

\begin{document}
\begin{flushright}
Accepted from publication in Physics Letters B\\
UM-P 014-2000\\
RCHEP 002-2000
\end{flushright}
\begin{center}
{\huge A note on low scale unification and gamma-gamma scattering}\\
\hspace{10pt}\\
S.R.Choudhury\footnote{src@ducos.ernet.in} \\
{\em Department of Physics, Delhi University, Delhi, India},\\ 
A. Cornell\footnote{a.cornell@tauon.ph.unimelb.edu.au} 
and G.C. Joshi\footnote{joshi@physics.unimelb.edu.au}\\
{\em School of Physics, University of Melbourne,}\\ 
{\em Parkville, Victoria 3108, Australia}\\
\hspace{10pt}\\
$31^{st}$ of October, 1999
\end{center}
\hspace{10pt}\\
\begin{abstract}
In this note we study an interesting effect of low energy gravity on 
photon-photon scattering at high energies.
\end{abstract}

\indent In a recent paper Gounaris, Porfyriadis and Renard \cite{one} have 
highlighted the possibility of exploring new physics through the scattering 
process $\gamma \gamma \rightarrow \gamma\gamma$ at c.m. energies in the 
TeV range.  The $\gamma\gamma$ mode is a possible mode of operation in a 
$e^+ e^-$ linear collider and this makes the study of this particular 
scattering at TeV-scale useful in a very practical way and not just of 
purely theoretical interest. Recently a lot of interest has also been 
generated in a TeV-scale unification wherin one envisages a Kaluza-Klein 
(KK) scenario in (4+n)-dimensions \cite{two,three}. Ordinary matter and 
gauge fields 
are localized in a (3+1)-dimensional brane configuration wheras gravity 
propagates in (4+n)-dimension, with the n-extra dimension compactified.  
For  n=2, the compactification scale turns out to be in the mm range with
a weak Planck scale in the TeV range (which acts as a cut-off for all
effective 
theories) and these numbers make this particular choice of n to be 
particularly interesting from the point of view of experiments in the 
collider.\\
\indent In the scenario just outlined excitations of the gravitons in the 
compactified dimensions would appear in the (3+1)-dimensional world as 
towers of particles.  At every level, there are one spin-2 state (massive), 
one spin-1 state and one spin-0 state.  The spin-1 state decouple from 
ordinary matter \cite{two,three} and the spin-0 state couples through the 
dilaton mode.  The spin-2 KK-modes with masses starting from the 1/R scale 
and effectively cutoff at  $M_s$ (that is at some scale of the order of 1 
or 2 TeV) couples to fermions and also to gauge particles and are the most 
visible signature of theories with compactification at low scale of the 
order of mms. The coupling of an individual KK-state is not of much 
interest since it is gravitationally suppressed, their interaction once 
summed over the towers of states however is significant.  It gets an 
effective strength that can be phenomenologically relevant to processes in 
the TeV-scale.  This note is to estimate such effects for $\gamma\gamma$ 
scattering at the TeV scale.  We show that the inclusion of the Spin-2 
KK-excitations of gravitons in the TeV-range results in changes in the 
scattering amplitude that is definitely within the measurement range of 
experiments in the colliders.\\
Consider the scattering process:\\
$$     \gamma(k_1,\lambda_1) + \gamma(k_2,\lambda_2) \rightarrow
     \gamma(k_3,\lambda_3)  + \gamma(k_4,\lambda_4)$$\\
where the k's and the $\lambda$'s refer respectively to momenta and 
helicities of the particles in the c.m. frame. The helicities take on 
values +1 and -1 . The invariant scattering amplitude for this process 
is denoted by  $ F_{h_1h_2h_3h_4}(s,t,u)$ where the h's take on values of 
signatures of the helicities and s,t are the usual Mandelstam variables.  
These amplitudes in the standard Model (SM) have been calculated by Jikia 
and Tkabladze \cite{four}. At values of s and t such that $s >> -t >> M_W^2$, 
the amplitudes  $F_{++++}$ and $F_{+-+-}$ together with their parity 
equivalents dominate:
\begin{equation}
 F_{++++}(s,t,u) = F_{+-+-}(s,t,u) 
\end{equation}
\begin{equation}
F_{++++}(s,t,u)=(-16\alpha^2i\pi)(s/t)\log (-t/M_W^2)
\end{equation}
Let us now estimate the contributions coming from KK-excitations in all 
the three channels. The vertex connecting the KK-excitations with a 
pair of photons have been explicitly worked out by Han, Lykken and 
Zhang \cite{three}. The resultant amplitdues have the symmetry:
\begin{equation}
F_{++++}(s,t,u)=  F_{+-+-}(u,t,s) 
\end{equation}
\begin{equation}
F_{----}(s,t,u) = F_{+-+-}(u,t,s)
\end{equation}
\begin{equation}
F_{-+-+}(s,t,u) =  F_{+-+-}(s,u,t) 
\end{equation}
and\\
\begin{equation}
F_{+--+}(s,t,u)  = (-i/4)* (\kappa^2)*(u^2)*[D(s)+D(t)]
\end{equation}
\begin{equation}
F_{-++-}(s,t,u)=F_{+--+}(s,t,u)
\end{equation}
where\\
\begin{equation}
D(s)=\sum \left( \frac{1}{s-M_{KK}^2} \right)
\end{equation}
$M_{KK}$ denoting the mass of the KK-excitations and the summation is over 
the entire tower of excitations. In the last equation, $M_{KK}$ is 
understood to include the width -i$\Gamma$/2 also.\\
\indent The contribution of these KK-excitations are complex in general.  
However, in respect of SM contributions, Gounaris et. al. \cite{one} have made 
the very important observation that the SM contribution is dominantly imaginary 
for values of s much greater than $M_w^2$ for directions away from 
the forward.  When the contributions coming from the KK-exchanges enter 
as corrections to the main SM contributions, clearly only the imaginary 
parts of the contributions become relevant.  The important point to note is 
that in the expression for the amplitudes only the s-channel resonance 
contributes an imaginary part whereas the others are completely real.  
Thus, in the range of energies where the KK-contributions are expected to 
be in the nature of correction terms to the main SM contribution, only 
the s-channel resonance contributions need be taken into account.  
This means only $F_{+-+-}$ and $F_{+--+}$ together with their parity 
equivalents need be considered. The imaginary parts of the KK-resonance 
contributions come from the imaginary parts of the propagator denominators:  
$(s-M_{KK}^2)$.  This is easily estimated following \cite{three}. The 
imaginary 
parts of the amplitudes above are given by:
\begin{equation}
\sum Im\left( \frac{1}{s-M_{KK}^2}\right) = 
-\frac{\pi s^{n/2-1} R^n}{\Gamma(n/2)(4\pi)^{n/2}}
\end{equation}
The nonvanishing contributions of the s-channel tower of KK-excitations to 
the imaginary parts of the amplitudes are thus given by:
\begin{equation}
ImF_{+-+-}= \left(\frac{\kappa^2 (1+\cos\theta_s)^2}{2}\right)
\left(\frac{\pi s^{n/2+1} R^n}{\Gamma(n/2)(4\pi)^{n/2}}\right)
\end{equation}
where $\theta_s$ is the c.m. scattering angle.  Im$F_{+--+}$is given by the 
above expression with cos$\theta_s$ replaced by its negative. Using now the 
relation (equation 64 of \cite{three})
\begin{equation}
\kappa^2 R^n = M_s^{-(n+2)} (4\pi)^{n/2} \Gamma(n/2)
\end{equation}
we get
\begin{equation}
Im F_{+-+-} = \left(\frac{\pi (1+\cos\theta_s)^2}{16}\right) (s^{1/2}/M_s)^4
\end{equation}
for n=2.\\

\begin{figure}
\includegraphics[width=10cm]{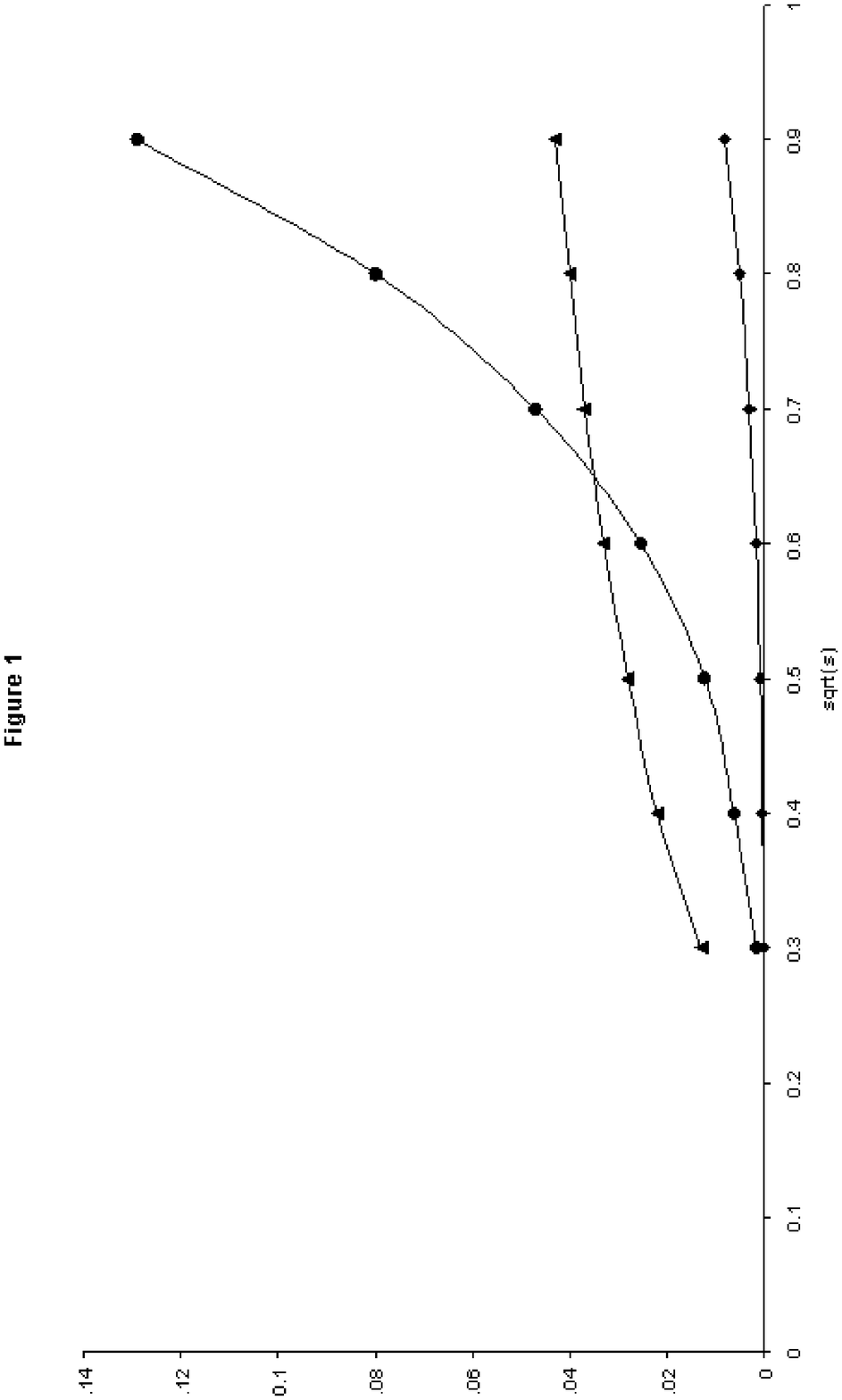}
\caption{Magnitudes of the imaginary parts of the dominant SM amplitudes 
and the contributions coming from s-channel KK-excitations. The 
contributions for $\theta_s=90^o$ are represented by triangles for the SM 
contributions, by circles for $F_{+-+-}$ with $M_s = 1TeV$, and by diamonds 
for $F_{+-+-}$ with $M_s = 2TeV$.  In this scenario, the amplitude $F_{+--+}$ 
equals $F_{+-+-}$.}
\label{fig1}
\end{figure}
\begin{figure}
\includegraphics[width=10cm]{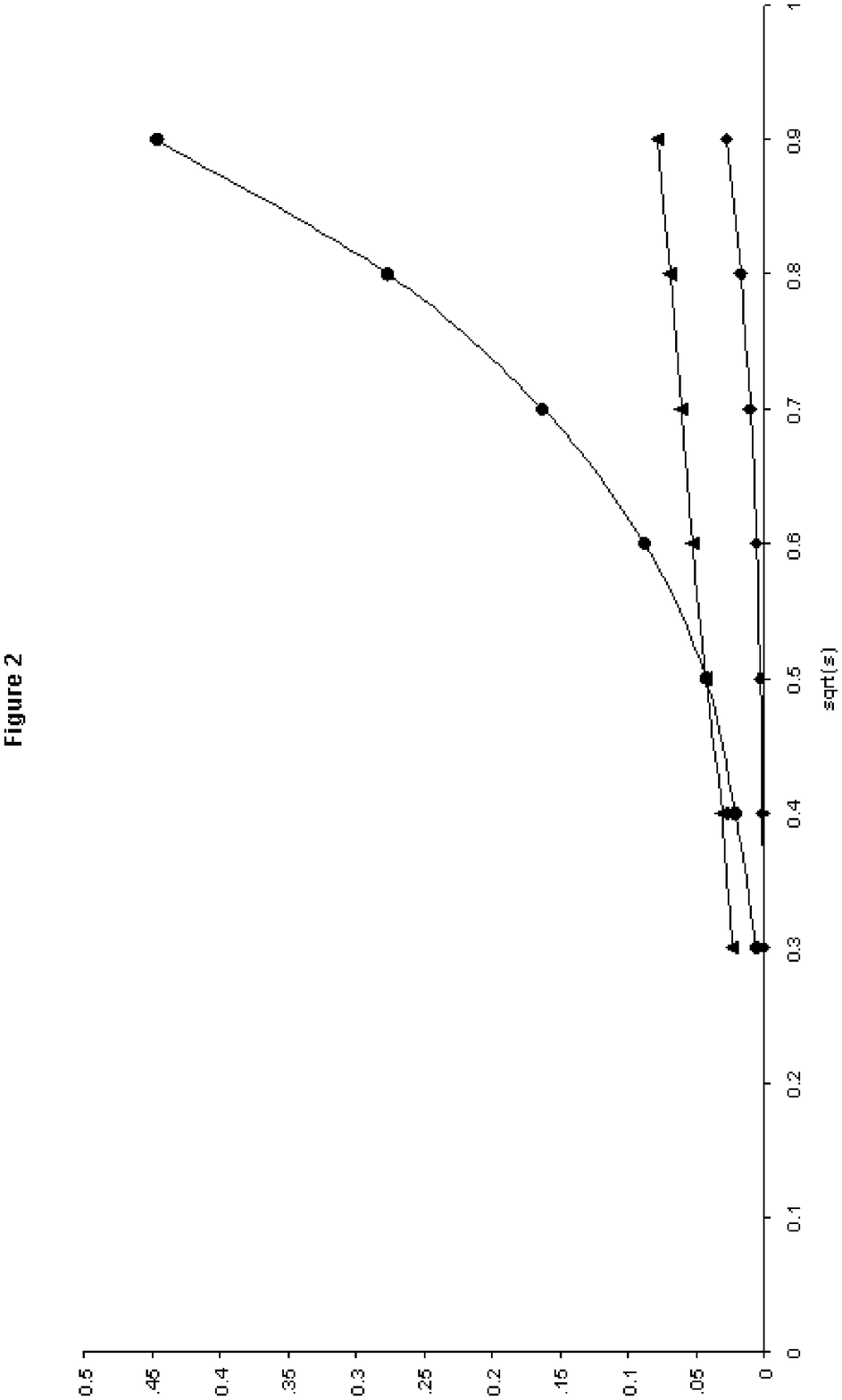}
\caption{Magnitudes of the imaginary parts of the dominant SM amplitudes 
and the contributions coming from s-channel KK-excitations. The contributions 
for $\theta_s=30^o$ are represented by triangles for the SM contributions, by 
circles for $F_{+-+-}$ with $M_s = 1TeV$, and by diamonds for $F_{+-+-}$ with 
$M_s = 2TeV$.  In this scenario, the amplitude $F_{+--+}$ is negligible.}
\label{fig2}
\end{figure}

\indent Multiple KK-excitation exchanges will give contributions proportional 
to higher powers of $s^{1/2}/M_s$ . These cannot be computed in a 
straightforward manner and thus the single KK-exchange contribution is a 
reliable correction only in the domain where $s^{1/2}/M_s$ is smaller than 
one.  We exhibit in figure 1 and 2, the relative contributions to the
imaginary parts of the amplitude $F_{+-+-}$ and $F_{+--+}$ in comparison to the 
SM predictions.  All other SM amplitudes are negligible in this energy and 
angle values and as reasoned above, the magnitudes of the KK-contributions 
can be taken seriously when  $s^{1/2}/M_s$ is not too close to unity.  
It is clear that there will be a window, whose value will depend upon the 
value of$M_s$, wherein the KK-exchange contributions will act as a correction 
term to the SM-dominant term with a definite magnitude.  Deviations of the 
measured cross-sections from the SM can thus be fitted to the correction 
term with a single parameter $M_s$ in the TeV. range. As the energies become 
higher so that multiple KK-exchange contributions become important as well, 
we are unable to calculate the KK-exchange contribution beyond saying that 
it will dominate the cross-section.  We have calculated above the 
cross-sections 
away from the forward direction.  Qualitatively similar conclusions of 
course can be drawn for the forward amplitude and hence for the total 
$\gamma$-$\gamma$ cross-section also.\\

\indent In conclusion, a low scale scenario leads to some definite pattern of 
deviation from the SM prediction for $\gamma\gamma$ scattering.  There exists 
a window at around a few hundred GeV. c.m. energy where the new physics gives 
rise a calculable correction to the SM values and thus provides a well 
defined signature.  At still higher energies, the contributions coming from 
multiple KK-exchanges start dominating over the SM but do not lend themselves 
to estimation in any reliable manner.\\\\ 
After this work was completed the following related papers on photon-photon 
scattering appeared \cite{five}.  However, our note emphasizes the
phenominalogical importance of a window around the few 100GeV.


\begin{thebibliography}{1}
\bibitem{one} G.J.Gounaris, P.I.Porfyriadis and F.M.Renard, 
	{\bf hep-ph 9812378}.
\bibitem{two} N.Arkani-Hamed, S.Dimopoulos and G.Dvali, 
	Phys.Lett. {\bf B429}, 263 (1998); 
	I.Antoniadis, N.Arkani-Hamed,S.Dimopoulos and G.Dvali, 
	Phys. Lett. {\bf B436}, 257 (1998); 
	N.Arkani-Hamed, S.Dimopoulos and J.March-Russel, 
	{\bf hep-th/9809124}.
\bibitem{three} Tao Han, J.D.Lykken, R.-J Zhang, {\bf hep-ph 9811350}.
\bibitem{four} G.Jikia and A.Tkabladze, Phys. Lett., 
	{\bf B323} (1994), 453.
\bibitem{five} O.J.P.Zboli, T.Han, M.D.Majso and P.G.Mercadante, 
	{\bf hep-ph/9908358}; H.Dovoudiasl {\bf hep-ph/9904425}; 
	K.Cheung {\bf hep-ph/9904266}.
\end{thebibliography}
\end{document}